**Title:** Follicle-stimulating hormone receptor: advances and remaining challenges.

**Authors:** Francesco De Pascali, Aurélie Tréfier, Flavie Landomiel, Véronique Bozon, Gilles Bruneau, Romain Yvinec, Anne Poupon, Pascale Crépieux, Eric Reiter*

**Affiliation:** PRC, INRA, CNRS, IFCE, Université de Tours, 37380, Nouzilly, France.

**\* Correspondence**: Institut National de la Recherche Agronomique (INRA) UMR85, CNRS-Université François-Rabelais UMR7247, Physiologie de la Reproduction et des Comportements - Nouzilly 37380, France - Email: *Eric.Reiter@inra.fr*

**Abstract**

Follicle-stimulating hormone (FSH) is produced in the pituitary and is essential for reproduction. It specifically binds to a membrane receptor (FSHR) expressed in somatic cells of the gonads. The FSH/FSHR system presents many peculiarities compared to classical G protein-coupled receptors (GPCRs). FSH is a large naturally heterogeneous heterodimeric glycoprotein. The FSHR is characterized by a very large NH2-terminal extracellular domain, which binds the FSH and participates to the activation/inactivation switch of the receptor. Once activated, the FSHR couples to G$\alpha$s and, in some instances, to other G$\alpha$ subunits. G protein-coupled receptor kinases and β-arrestins are also recruited to the FSHR and account for its desensitization, the control of its trafficking and its intracellular signalling. Of note, the FSHR internalization and recycling are very fast and involve very early endosomes instead of early endosomes. All the transduction mechanisms triggered upon FSH stimulation lead to the activation of a complex signalling network that controls gene expression by acting at multiple levels. The integration of these mechanisms leads to context-adapted responses from the target gonadal cells, but also indirectly affects the fate of germ cells. Depending of the physiological/developmental stage, FSH elicits proliferation, differentiation or apoptosis in order to maintain the homeostasis of the reproductive system. Pharmacological tools targeting FSHR recently came to the fore and open promising prospects both for basic research and therapeutic applications. This paper provides an updated review of the most salient aspects and peculiarities of FSHR biology and pharmacology.

# 1. Introduction.

Follicle-stimulating hormone (FSH) plays a crucial role in the control of male and female reproduction. FSH is a heterodimeric glycoprotein consisting of an α subunit, common with the other glycoprotein hormones [i.e.: luteinizing hormone (LH), chorionic gonadotropin (CG) and thyroid-stimulating hormone (TSH)], which is non-covalently associated with a specific FSHβ subunit (Pierce and Parsons, 1981, Ryan et al., 1987). FSH is synthesized and secreted by the pituitary. FSH binds to and activates a plasma membrane receptor (FSHR) that belongs to the rhodopsin family of the G protein-coupled receptor (GPCR) superfamily. The FSHR displays a high degree of tissue specificity, being expressed in Sertoli and granulosa cells located in the male and female gonads respectively (Simoni et al., 1997) (**Figure 1**). As the other glycoprotein hormone receptors, the FSHR is characterized by a large NH2-terminal extracellular domain (ECD), where FSH binds specifically.

FSH is required for normal growth and maturation of ovarian follicles in women and for normal spermatogenesis in men (Themmen and Huhtaniemi, 2000). Female mice with FSHβ or FSHR gene knockout are infertile because of an incomplete follicle development, whereas male display oligozoospermia and subfertility (Kumar et al., 1997, Dierich et al., 1998). Consistently, women expressing non-functional variants of the FSHR are infertile while men are oligozoospermic, yet fertile (Aittomäki et al., 1995).

Because of its glycosylation, FSH is naturally heterogeneous and must be expressed by mammalian cells (*i.e.*: pituitary or CHO cells) to be fully active *in vivo*. Because of these characteristics, only native forms of FSH, either purified from urine or recombinant, are being used in reproductive medicine, no other pharmacological agents being currently available in clinic (Lunenfeld, 2004, Macklon et al., 2006). Some women treated with FSH develop an ovarian hyperstimulation syndrome (OHSS), which, in its severe forms, can be life-threatening (Vloeberghs et al., 2009). Therefore, pharmacological agents that would induce ovulation without the risk of provoking OHSS would represent a major improvement. It is also well established that, in women, the responsiveness to FSH treatment is heterogeneous and that the dose and sometimes the source of FSH, have to be empirically adjusted for each patient (Loutradis et al., 2003, Loutradis et al., 2004). A larger panel of FSHR agonists with varying pharmacological profiles could certainly help improving the overall efficiency of medically-assisted procreation. On the other hand, FSHR blockers could potentially represent a novel non-steroidal approach for contraception (Naz et al., 2005).

In order to meet these challenges, it is important to gain a better understanding of FSHR biology and the bottlenecks that make the targeting of this receptor particularly difficult.

## 2. FSH and FSH-R in pathologies.

FSH serum levels vary physiologically during the menstrual cycle in women. Nevertheless, abnormal pituitary FSH secretion can occur in different pathologies such as Premature Ovarian Insufficiency (POI) and Polycystic Ovarian Syndrome (PCOS). POI is a dysfunction of the ovary occurring in about 1% of female population (under 40 years old)(Goswami and Conway, 2005). Patients carrying POI are infertile due to anovulation, amenorrhea and reduced secretion of oestrogens (Kalantaridou et al., 1998). One key parameter for the diagnosis is based on elevated FSH serum levels (> 40 IU/l) found in non-menopausal women (Conway, 2000). The higher FSH levels are caused by diminished oestrogens production that abolishes the negative feedback at the pituitary gland resulting in hypersecretion of FSH and LH (hypergonadotrophic hypogonadism). Interestingly, a woman with a mutation in the β-subunit of FSH presented infertility and amenorrhea. This mutation resulted in a truncated β-subunit, which was unable to bind the α-subunit and to activate the receptor (Matthews et al., 1993). These authors hypothesized that mutation at the FSH level could represent a possible cause of POI but a parallel study was unable to find any mutation in the hormone in a group of 18 women affected by POI (Layman et al., 1993). Another worldwide women–affecting disease (5-10%) (Franks, 2013), in which FSH and its receptor seem to play a role, is PCOS. PCOS is the commonest cause of infertility due to anovulation in women (Ben-Shlomo and Younis, 2014) and, according to Rotterdam criteria, the diagnosis is based on the presence of oligomenorrhoea, hyperandrogenism and ovary with diffuse cysts (Rotterdam, 2004). The causes are not fully understood but many studies demonstrate environmental and genetic (Ben-Shlomo and Younis, 2014) contributions as well as a specific evolutionary cause (Casarini and Brigante, 2014, Casarini et al., 2016d). Since low levels of serum FSH are encountered in PCOS, an open debate regards the possible correlation of FSH levels as a cause for the anovulation found in PCOS. In this scenario, the interpretation of anovulation denotes the incapacity of FSH to select the dominant follicles and lead them to maturation, consequently, there is an accumulation of immature follicles in the ovary (Franks et al., 2000). This possible implication of FSH hypo-responsiveness is in disagreement with *in vitro* experiments on granulosa cells derived from anovulatory women suffering of PCOS, who showed a hypersensitivity to FSH stimulation in terms of oestradiol production. This apparent paradox could be clarified by considering the abnormal endocrine environment, together with the heterogeneity of the follicles (different follicle sizes have different responses to gonadotropins) (Willis et al., 1998) found in ovaries of women affected in PCOS-derived anovulation (Franks et al., 2008). Genetically, the involvement of two common FSHR polymorphisms (Thr307Ala and Asn680Ser) has been considered in the pathogenesis of PCOS but no stronger and definitive correlation has been demonstrated to date (Chen et al., 2014). In male, FSH/FSHR contribution to infertility is not fully elucidated due to a lack of knowledge about the role of FSH in spermatogenesis. In the clinical practice, administration of FSH is used to treat idiopathic male infertility (IMI) (Ulloa-Aguirre and Lira-Albarrán, 2016) but this treatment remains controversial since disrupting mutations in the FSH/FSHR has no effect on fertility in human and mouse (Kaprara and Huhtaniemi, 2017). Finally, a pathogenic function has been recently assigned to the FSH/FSHR system in tumours. Some studies show the expression of FSHR in different types of tumours including ovarian, prostate and endothelial cancers, but the data are debated and based on small cohort of patients (Papadimitriou et al., 2016). Further investigations is necessary to clarify the role of FSH/FSHR in malignancies.

### 3. FSH role in assisted reproduction technologies.

Assisted Reproduction Techniques (ART) are defined as "all treatments or procedures that include the *in vitro* handling of both human oocytes and sperm or embryos for the purpose of establishing a pregnancy" (Zegers-Hochschild et al., 2009). The procedures involved in ART include *In Vitro* Fertilization (IVF), Gamete Intrafalloppian Transfer (GIFT) and Intracytoplasmic Sperm Injection (ICSI). Although a universal ART protocol does not exist, the main steps are common to each technique (Casarini et al., 2016a). Briefly, the starting point is the suppression of the natural menstrual cycle and pituitary gland function, which is followed by the induction of ovulation through the injection of urinary or recombinant FSH preparations in a procedure called Controlled Ovarian Stimulation (COS). After the follicle has reached a proper size, maturation of the oocyte is obtained by administration of drugs, a phase known as ovulation triggering. The following step is the selection and collection of the best oocytes that will ultimately undergo fertilization by using one of the aforementioned techniques. Finally, the fertilized oocyte is re-implanted in the uterus and the luteal phase is pharmacologically supported (Farquhar et al., 2015). The ovarian stimulation is the main responsible for the success or failure of ART. The purpose is to obtain as many as possible growing follicles, in order to allow oocytes to mature and consequently to be retrieved and fertilized. This goal is achieved by maintaining FSH blood serum levels above the threshold, simply subjecting women to daily FSH injections (Baird, 1987). The choice between urinary or recombinant FSH is completely arbitrary since different studies have demonstrated the equivalence of the two preparations in terms of efficacy and safety (Daya and Gunby, 2006). Since a standard protocol for ovarian stimulation is not available, clinicians adjust FSH dose administration in a patient-personalized mode, taking into account the different degrees of ovarian responsiveness to FSH. In this context, women undergoing COS protocol for assisted reproduction are generally divided in normal, poor and high responders. The range of urinary or recombinant FSH administration goes from 150 IU for high responders up to 375 IU for poor responders whereas a medium concentration (200 – 300 IU) is used for normal responders (Alviggi et al., 2012). It has been suggested that poor responders might benefit of the addition of luteinizing hormone (LH) in the FSH preparation but more solid demonstrations are needed (Hill et al., 2012). The CONSORT study attempted to create an algorithm to calculate the personalized dose of recombinant FSH in ART. The algorithm considers different parameters (basal FSH serum levels, body mass index, age, etc.) and proposes a mathematical model for FSH dose calculation that can be used as a starting point for personalized FSH administration in ART protocols (Olivennes et al., 2009). A severe complication due to gonadotropin administration in ART is the Ovarian Hyperstimulation Syndrome (OHSS). The incidence is variable but it can be as high as 20% (Smith et al., 2015) in women treated for ART. OHSS are classified as mild, moderate or severe and result in exaggerated ovarian response to gonadotropins. The causes are still unknown but many studies attempted to correlate OHSS with different parameters (antral follicle count, anti-Müllerian hormone levels, oestradiol concentration, etc.) in order to find risk factors and consequently delineate preventive strategies (Nastri et al., 2015). Mutations at the FSH receptor were also considered as a risk factor for OHSS occurrence but the mutations were found in a spontaneous form of OHSS (Dieterich et al., 2010). Infact, FSH receptor mutations are generally very rare in ART-induced OHSS. On the contrary, FSHR single nucleotide polymorphisms (SNPs) Ser680Asn correlates with a high severity of OHSS in women carrying this SNP (Rizk, 2009). Finally, genetics of FSH and FSHR, including mutations, SNPs or combination of them resulting in different haplotypes might affect COS and consequently assisted reproduction outcome, or might be used as genetic marker of ovarian response, but further supporting evidences is needed (Riccetti et al., 2017).

## 4. FSHR structure and function.

The human FSHR, together with LHR and TSHR, belongs to the glycoprotein hormone receptors subfamily of class A GPCRs. It is encoded by a unique gene constituted of 10 exons and located on chromosome 2p21-p16 (Rousseau-Merck et al., 1993). After releasing of a signal peptide of 17 amino acids, the mature membrane FSHR protein contains 678 amino acids. Its molecular weight varies between 82 and 89 kDa depending on the rate of N-glycosylation (Davis et al., 1995). Several splice variants have been reported in human, [for a review (Ulloa-Aguirre and Zarinan, 2016)], some of them coding for altered isoforms with no or reduced signalling potential (Gerasimova et al., 2010, Karakaya et al., 2014, Zhou et al., 2013).

The FSHR structure consists in an extracellular hydrophilic domain (ECD) composed of a long N-terminal part and three loops. This ECD connects to a hydrophobic region comprising seven transmembrane domains (TMDs) and intracellular hydrophilic regions composed of three intracellular loops and the C-terminal part (**Figure 2**) (Jiang et al., 2012, Jiang et al., 2014a, Fan and Hendrickson, 2005, Ulloa-Aguirre and Zarinan, 2016). The ECD can be functionally divided into two subdomains, the hormone binding domain (HBD) consisting of ten consecutive leucine-rich repeats domains (LRR) and the hinge region located between the HBD and the first TMD. The hinge region contains two additional LRR, followed by a hairpin loop and an alpha helix flanked on both sides by two cysteine box motifs that are able to form cysteine bonds between themselves. The hinge ends with a highly conserved decapeptide (FNPCEDIMGY), located close to the first TMD reported to be necessary for receptor activation (Bruser et al., 2016). A currently proposed model predicts that ligand-induced activation of FSHR operates following a two-step process. The first step corresponds to the binding of FSH on the HBD. The crystal structure of the hormone in complex with the HBD has been solved (Fan and Hendrickson, 2005). The winding structure determined by the ten LRR requires that the HBD takes the form of a curved tube. The dimensional difference between the inner and outer surface of this tubular structure is responsible for the curvature (Jiang et al., 2012, Jiang et al., 2014a). Analysis of the FSH-HBD crystal shows that FSH binds into the concave face of the curved HBD domain where the beta sheets of the LRR motifs are located. The long axis of FSH is perpendicular to the HBD tube and aligned with the LLR beta sheets. Therefore, FSH associates with the HBD like a « handclasp » (Jiang et al., 1995). The buried interface has a high charge density. Contact with hormone is mainly ensured by ten residues, charged for the most part (L55, E76, D81, R101, K104, Y124, D150, D153, K179, I222). Some of them participate to the FSHR specificity for FSH (L55, E76, R101, K179, I222) and discrimination against LH/CG (L55, K179) and TSH (E76, R101) binding [reviewed in (Jiang et al., 2014a)]. The carbohydrates do not take part in the binding interface (Fan and Hendrickson, 2005). Nevertheless, two of the three possible N-glycosylation sites of the ECD can actually be glycosylated (N174 and N276) and at least one of the two is required for binding FSH activity (Davis et al., 1995).

The binding of FSH on the HBD enables the second step of the ligand-induced receptor activation consisting of a conformation change in the hinge region. The crystal structure of the entire FSHR ECD bound to FSH has highlighted the role of the hinge region for FSHR activation (Jiang et al., 2012, Jiang et al., 2014a). FSH binds more tightly the HBD when the hinge region is present than in its absence. Even if the entire ECD shows a greater accessible buried interface than the HBD alone, leading to a more rigid FSH conformation, the main reason for the gain in affinity comes from the presence of a sulfated tyrosine (Y335) in the hairpin loop of the hinge that constitutes a second interaction site between FSH and its receptor. FSH binding to the inner surface of the HBD induces the formation of a sulfated-tyrosine pocket between the FSH $\alpha$ and $\beta$ chains, absent in the free form of the hormone. The sulfated tyrosine-binding pocket interacts with the sulfated-Y335 and neighbouring residues of FSHR and hence provokes the tilting of the hairpin loop and the rotation of the $\alpha$-helix of the hinge region. This causes a conformational change of the TMDs, then the extracellular loops may interact with the charged residues in the hairpin loop and finally stabilize the active state of the receptor. It has also been proposed that the hinge region could act as a tethered inverse agonist keeping the RFSH in its inactive conformation. The

FSH binding on the HDB could alleviate this inhibitory effect by changing the receptor conformation. Consistent with this view, the removal of the hinge region, with the exception of the conserved decapeptide, leads to increased constitutive activation level of FSHR and loss of its ability to respond to FSH (Agrawal and Dighe, 2009). The inactive state driven by the hinge region and maintained by an unoccupied HDB, coupled with a discriminating binding interface for LH/CG and TSH, allows the FSHR to remain inactive in the absence of FSH, even when high concentrations of others glycoproteins are circulating. Constitutively active mutants of FSHR are rare, probably because they induce severe pathologies (Ulloa-Aguirre et al., 2014). The mutations responsible for three of them target two different TMDs and induce conformational changes of FSHR that decrease its binding specificity. Consequently, these mutant forms of FSHR can be activated by high concentrations of CG or TSH.

Like others glycoprotein hormone receptors, FSHR was observed as a monomer/oligomer. It has been shown, using FRET microscopy, that FSHR can form oligomers in endoplasmic reticulum early during biosynthesis (Thomas et al., 2007). In this study, oligomers of FSHR were observed at the cell surface and FSH had little effect on FSHR oligomerization. Later on, it was shown that FSHR could heterodimerize with LHR (Mazurkiewicz et al., 2015). Fan and Hendrickson showed that the dimerization of the ectodomain occurs only through a weak interaction which could be mediated via residue Y110, suggesting that a stronger interaction could be carried out by another domain of the receptor, notably by the alpha-helices of the TMDs, as suggested by Ulloa-Aguirre and Zarinan (Ulloa-Aguirre and Zarinan, 2016, Fan and Hendrickson, 2005). A model for a trimeric FSH receptor has also been proposed by Jiang et al. (Jiang et al., 2012, Jiang et al., 2014a, Jiang et al., 2014b). According to this model, only one fully glycosylated FSH molecule would be able to bind to a trimer and therefore just one of the three receptors would be activated.

5. **FSHR mutations.**

Although rare in the population, activating and inactivating mutations of FSHR have been reported in both genders (Riccetti et al., 2017, Desai et al., 2013, Ulloa-Aguirre et al., 2014). Both types of mutations can cause alteration of reproductive function, even though the phenotype is often more severe for woman fertility. In most cases, inactivating mutations provoke primary or secondary amenorrhea whereas the activating ones generally lead to an OHSS. Some studies described the impact of mutations on both the functional properties of the mutated receptor and the resulting phenotype and contributed to extend the knowledge of the structure/function relationship of FSHR.

Inactive mutations in the ECD mostly impair cell surface expression/trafficking of the receptor while FSH binding of the hormone is often reduced. Accordingly, signal transduction is negatively impacted, leading to a greater or lesser decrease of the production of FSH-induced cAMP. However, Aittomaki et al. showed that the A189V mutation, located in the HDB, provokes the decrease of the cAMP responsiveness induced by FSH, but has no effect on the affinity of the FSHR for the hormone (Aittomäki et al., 1995). Some residues in the HDB may play a different role than those merely involved in binding affinity for the ligand. Similarly, FSHR carrying the P348R mutation in the hinge region does not bind FSH, suggesting that residues outside of the HDB participate to FSH binding (Allen et al., 2003).

In TMD2, two different mutations lead to a decrease by half of the production of FSH-induced cAMP. Bramble et al., have reported that the D408Y mutation impairs the trafficking of the receptor, the decrease of both the expression at the cell surface and the signal transduction being proportional (Bramble et al., 2016). For the A419T mutation, the cell surface expression was not evaluated, but the ligand binding capacities of the mutated and wild type receptors were similar (Doherty et al., 2002).

Two mutations located in TMD6 present different impact on signal transduction. The A575V mutation leads to an increase of cell surface expression but a slight decrease of the binding affinity, and therefore does not show any difference in FSH-induced cAMP compared to wild type (Desai et al., 2015). Kuechler et al. showed that the P587H mutation completely abolished cAMP production upon FSH stimulation (Kuechler et al., 2010). This mutation could induce a distortion of a proline-dependent helix-kink at this position, which is particularly important for signal transduction in class A GPCRs. The P519T mutation located in extracellular loop 2 (ECL2) decreases FSH binding and cell surface expression, impairing cAMP response *in vitro*. This mutated receptor could remain trapped intracellularly due to the alteration of the cell surface targeting. The decrease of ligand affinity suggests that ECL2 is involved in the binding mechanism, possibly *via* direct interaction with the ECD (Meduri et al., 2003). In ECL3, mutation of the 601 residue impairs the signal transduction without reducing the binding capacity although two diverging reports were published. Ryu et al., reported that mutation L601A improves FSH binding while Touraine et al. showed that this mutation does not affect FSH binding (Ryu et al., 1998, Touraine et al., 1999). The cellular trafficking of the L601V mutant is not affected and does not explain the dramatic decrease of FSH-induced cAMP response. Similarly, R573C mutation also located in the ICL3, decreases signal transduction without reducing hormone binding capacity and cell surface expression (Beau et al., 1998). Hugon-Rodin et al. first described a mutation in the intracellular C-terminal tail of RFSH in the case of a non-pregnant women with OHSS (Hugon-Rodin et al., 2017). The R634H mutation is located in the highly conserved BXXBB motif (B represents a basic residue and X any non-basic residue), which is needed for cell surface trafficking and plasma membrane expression (Ulloa-Aguirre et al., 2007, Duvernay et al., 2004). This mutant displays decreased cell surface expression that leads to a lowered cAMP production in response to FSH. However, as this mutation was identified in an a OHSS patient, an activating mutation was expected. Moreover, this mutant does not exhibit any constitutive activity. Further investigation will be necessary to solve this conundrum and find a causal link between the mutation and the clinical data.

Activating mutations of FSHR have been identified in most structural/functional domains of the FSHR and have been shown to lead to OHSS by different molecular mechanisms. Some mutated receptors display increased constitutive activity and promiscuous binding to hCG and TSH compared to wild type FSHR: T449N and T449A in TMD3 (Banerjee et al., 2017, Montanelli et al., 2004), I545T in ID3 (Montanelli et al., 2004, Smits et al., 2003) and I545T in TMD5 (De Leener et al., 2008). De Leneer et al suggested that the gain of sensitivity of the mutants to hCG (or TSH) could be due to the lowering of an intramolecular barrier aiming at preventing activation rather than to an increase in binding affinity (De Leener et al., 2008). These mutated receptors respond to the increasing hCG levels occuring during pregnancy and TSH levels in case of hypothyroidism. Activating mutations displaying other properties have also been reported. De Leneer et al. described the S128T mutation located in the ECD (De Leener et al., 2008). This mutant displays an increased sensibility for hCG due to a greater affinity but not for TSH, and does not present any constitutive activity. In this case, the mutation could modify the interaction between residues from hCG and HDB. Mutants devoid of constitutive activity or increased responsiveness to hCG and TSH have also been described. Desai et al. reported the V514A mutation located in the ECL2, which induces an increase of cell surface expression and binding affinity along with reduced internalization (Desai et al., 2015). Interestingly, this mutant displays better FSH responsiveness for cAMP production. Even though maximal efficacy does not seem to change, the improved signal transduction at low FSH concentration could explain the advent of OHSS in this patient.

### 6. FSHR signalling through G proteins.

Unlike many GPCRs, in the absence of ligand, FSHR displays little to no constitutive activity (Ulloa-Aguirre et al., 2014). This functional characteristic correlates with the increased stability of the transmembrane domains in the inactive state compared to other glycoprotein hormone receptors (Zhang et al., 2007, Ulloa-Aguirre et al., 2014). Upon FSH binding, conformational changes in the receptor lead to the transduction of extracellular signal, hence the activation of several intracellular signalling pathways dependent of heterotrimeric G proteins. In all animal species and in physiological conditions, the primary transduction mechanism described for FSHR involves the heterotrimeric Gαs proteins (Abou-Issa and

Reichert, 1979, Ayoub et al., 2015, Musnier et al., 2009, Dattatreyamurty et al., 1987, Remy et al., 1995, Gershengorn and Osman, 2001). The G$\alpha$s-subunit interacts with intracellular loops 2 and 3 of the activated FSHR at the level of the ERW and BBXXB (B: basic residue, X: non-basic residue) motifs respectively (Jiang et al., 2012, Ulloa-Aguirre and Zarinan, 2016, Ulloa-Aguirre et al., 2007). Several intracellular events are then observed: 1) exchange of G$\alpha$s-subunit-linked GDP to GTP, 2) dissociation of G$\alpha$s-subunit-GTP and G$\beta\gamma$ subunits, and 3) stimulation of the adenylate cyclase by G$\alpha$s-subunit-GTP inducing an increase of intracellular cAMP levels and its propagation into the cell. The main function of cAMP is to activate protein kinase A (PKA) that in turn phosphorylates various substrates such as transcription factor cAMP response element-binding protein (CREB)(Hunzicker-Dunn and Maizels, 2006, Hansson et al., 2000). Intracellular cAMP also leads to the activation of the exchange proteins directly activated by cAMP (EPACs) (Kawasaki et al., 1998, Wayne et al., 2007, de Rooij et al., 1998). It has been reported that the G$\alpha$s/cAMP/PKA pathway may be lost in some FSHR mutants derived from altered splice transcripts in exons 10 which encodes the G protein coupling domains (Sairam et al., 1996). Noteworthy, the cAMP pathways is strongly amplified downstream of FSHR since maximal response can be reached with only a few percent of FSHR being occupied (Ayoub et al., 2015). In line with many other GPCRs, FSHR has demonstrated a certain degree of promiscuity with other G proteins rather than being exclusively coupled to G$\alpha$s. Indeed, it has been reported that FSHR is able to interact with G$\alpha$i (Gorczynska et al., 1994) and G$\alpha$q (Quintana et al., 1994). FSHR is able to activate pertussis toxin-sensitive G$\alpha$i protein in presence of glycosylated variants of FSH (Arey et al., 1997, Timossi et al., 2000) and in Sertoli cells at specific stages of maturation (Crepieux et al., 2001) whereas it has been reported to activate G$\alpha$q protein when FSH reaches supra-physiologic concentrations (> 200 ng/ml) (Ito et al., 2003, Gloaguen et al., 2011, Conti, 2002).

## 7. FSHR desensitization, internalization and recycling.

The FSHR is regulated by the canonical desensitization mechanisms known to operate for most GPCR (**Figure 3**). Briefly, agonist-activated FSHR is rapidly phosphorylated on serine/threonine residues located on its carboxyl terminus through the action of G protein-coupled receptor kinases (GRKs), specifically GRKs 2, 3, 5 and 6 (Ayoub et al., 2015, Kara et al., 2006, Lazari et al., 1999, Marion et al., 2006, Moore et al., 2007, Troispoux et al., 1999, Ulloa-Aguirre and Zarinan, 2016). Non-visual arrestins (β-arrestin 1 and 2) are subsequently recruited and overlap with G protein binding sites on the receptor (Ayoub et al., 2015, Nakamura et al., 1998, Stoffel et al., 1997). Through this interaction, β-arrestins -1 and -2 prevent the coupling between FSHR and heterotrimeric G proteins by steric hindrance (Reiter et al., 2017, Reiter and Lefkowitz, 2006, Thomsen et al., 2016). The amount of intracellular cAMP then decreases and the receptor becomes refractory to further FSH stimulation (Marion et al., 2002). Once desensitized, the FSHR is thought to follow intracellular routing and molecular events similar to those that have been largely demonstrated for other GPCRs. In short, the receptor is internalized *via* the clathrin-dependent endocytic pathway, resensitized and recycled back at the plasma membrane. Upon binding to the FSHR, β-arrestins undergo conformational rearrangements inducing the exposition of binding motifs in the C-tail that interact with various elements of the endocytic and recycling machinery (Luttrell and Lefkowitz, 2002, Moore et al., 2007, Tian et al., 2014) such as clathrin (Goodman et al., 1997, Goodman et al., 1996), clathrin-associated adapter complex AP2 (Laporte et al., 2000, Laporte et al., 1999), phosphoinositides (PIP2, phosphatidylinositol 4,5-bisphosphate, IP3, phosphatidylinositol 3,4,5-trisphosphate and IP6, inositol hexakisphosphate) (Gaidarov et al., 1999, Toth et al., 2012), dynamin (Zhang et al., 1996), ADP-ribosylation factors (ARFs) specifically ARF-6 and ARNO its activity modulator (ARF nucleotide-binding site opener) (Frank et al., 1998, Shmuel et al., 2006); small G proteins (Rab) (Moore et al., 2007) and N-ethylmaleimide-sensitive fusion protein amine (NSF) (McDonald et al., 1999). Before recycling to the plasma membrane, the receptor localized in endosomal vesicle must be resensitized (Luttrell and Lefkowitz, 2002). The receptor is then dephosphorylated by phosphatase PP2A recruited by β-arrestins and dissociated from its agonist by an acidic pH in the vesicle (Luttrell and Lefkowitz, 2002, Thomsen et al., 2016). FSHR exhibits well the desensitization and internalization properties of class A GPCRs. Indeed, the internalization is transient, rapid and maximal between 2 and 5 min post-stimulation, the receptor resensitization is predominant compared to the degradation and the recycling process is fast (half time of about 10 min) (Ayoub et al., 2015, Moore et al., 2007, Thomsen et al., 2016, Kara et al., 2006). Unlike Gαs-coupling, more than 90% of FSHR occupancy is required in order to achieve maximal β-arrestin recruitment and internalization (Ayoub et al., 2015).

Early endosomes (EE) have been considered as the primary site for initial post-endocytic sorting of some GPCRs (Sposini and Hanyaloglu, 2017). However, FSHR has recently been reported to traffic to very early endosomes (VEE) for its post-endocytic sorting. The VEE is an endosomal compartment distinct from the EE, which plays a key role in direct spatial control of FSHR but also LHR and β1AR signalling (Jean-Alphonse et al., 2014). The VEE is smaller than EE and lacks typical EE markers. However, the adaptor protein containing PH domain, PTB domain, and Leucine zipper motif (APPL1) is expressed in VEE subpopulations (Jean-Alphonse et al., 2014, Sposini et al., 2017). Interestingly, the divergent sorting of FSHR between EE and VEE is mediated by association with the PDZ protein Gαi interacting protein C terminus (GIPC). Moreover, the sustained ERK activation profile elicited upon FSHR stimulation requires both internalization and targeting to VEE in a GIPC-dependent manner (Jean-Alphonse et al., 2014).

8. **FSHR signalling through β-arrestin.**

Beyond their well established role in receptor desensitization, internalization and recycling, β-arrestins have progressively emerged as key players in the control of GPCR-mediated signals in time and space (**Figure 3**). Many GPCRs, including FSHR, have been demonstrated to signal independently of heterotrimeric G protein, through ligand-induced β-arrestin 1 and 2 recruitment (Reiter et al., 2012, Reiter et al., 2017, Reiter and Lefkowitz, 2006). Indeed, β-arrestins act as multifunctional scaffolds and activators for a large number of signalling proteins (Xiao et al., 2007, Xiao et al., 2010, Crepieux et al., 2017). A well-documented illustration of the temporal encoding in GPCR-induced signalling pathways that β-arrestins can exert is the dual mechanism of ERK activation reported for several GPCRs including the FSHR. G protein-mediated ERK activation is rapid and transient. In contrast, the ERK signal activated via β-arrestins is slower in onset (~5-10 min to reach maximum), but protracted (t1/2 > 1 hour) (Ahn et al., 2004, Kara et al., 2006).

Importantly, β-arrestin-dependent signalling is a tightly regulated process. Upon FSH binding, the FSHR adopts an active conformation and becomes a target for GRK phosphorylation. Specifically, GRK5 and 6 have been reported to compete with GRK2 and 3 for the phosphorylation of different receptors including the FSHR. Indeed, GRK5 and 6 have been shown to promote β-arrestin-dependent ERK activation whereas GRK2 and GRK3 inhibit the same pathway (Kara et al., 2006, Heitzler et al., 2012, Kim et al., 2005, Ren et al., 2005, Shenoy et al., 2006). In line with these results, recent studies have provided direct evidence of the phosphorylation bar code for different GPCRs (Busillo et al., 2010, Butcher et al., 2011, Lau et al., 2011, Nobles et al., 2011). Collectively, these findings support a model in which the stabilization of distinct receptor conformations subsequently leads to specific phosphorylation patterns involving distinct GRK subtypes. This phosphorylation bar code could then control the conformation of β-arrestin recruited to the receptor and thereby its ability to interact with specific partners. This cascade of molecular events would ultimately control receptor fate and the β-arrestin-dependent intracellular signalling in a specific manner

(Reiter et al., 2012). It has recently been demonstrated that 5-HT2R, β2AR, CXCR4 and FSHR activate the ERK pathway *via* a mechanism involving MEK-dependent β-arrestin 2 phosphorylation at Thr383. Importantly, this agonist-induced phosphorylation of β-arrestin 2 is a necessary step for ERK recruitment to β-arrestin complex and for ERK activation (Cassier et al., 2017). In addition to ERK, β-arrestins have been reported to be required for the activation of p70S6K and rpS6 (Trefier et al., 2017, Wehbi et al., 2010b).

In the classical view, G protein signalling originates at the cell surface and is followed by rapid β-arrestin-mediated quenching of G protein signalling. Recent findings have begun to challenge this paradigm. A number of GPCRs have been reported to elicit sustained G protein signalling, rather than being desensitized after initial agonist stimulation (Calebiro et al., 2009, Feinstein et al., 2011, Ferrandon et al., 2009, Irannejad et al., 2013, Mullershausen et al., 2009). In line with these data, it has been proposed that, for some GPCRs, a series of distinct signalling waves could arise upon activation (Lohse and Calebiro, 2013): i) a first wave triggered at the cell surface upon G-protein coupling and second messenger release; ii) a second wave originating from clathrin-coated pits/vesicles where β-arrestin bound to the receptor would induce signals such as ERK activation and iii) a third wave involving signalling via G proteins from the endosomal compartment. X-ray crystallography of the β2AR in complex with Gαs has revealed that the interaction involves both the N-terminal and C-terminal domains of the Gαs subunit and the core of the receptor (*i.e.*: ICL2, TM5, and TM6) (Rasmussen et al., 2011). Importantly, a recent study revealed that β-arrestins interact with two different sites on the receptor; one is the phosphorylated receptor carboxyl terminus and a second, within the core of the receptor (Shukla et al., 2014). Moreover, internalized receptor complexes called "megaplexes" composed of a single GPCR, β-arrestin, and G protein were recently discovered and their architecture and functionality were described (Thomsen et al., 2016). Even though the validity of this model is not yet fully established in the case of the FSHR, recent works from Hanyaloglu's group on LHR and FSHR suggest that analogous mechanisms could operate for these receptors as well (Jean-Alphonse et al., 2014, Sposini et al., 2017).

### 9. Modelling of FSHR signalling.

FSH signalling acts at different time scales within the hypothalamic-pituitary-gonadal (HPG) axis, encoding and decoding complex signals across several organs and tissues from the pituitary cells to the somatic cells in the gonads. Capturing the mechanisms responsible for such refined controls has proven very challenging. Over the years, this topic has led to the development of numerous mathematical models.

Fluctuations of FSH circulating levels are tightly regulated with respect to those of LH, GnRH and steroids, displaying striking temporal patterns. A large part of mathematical modelling of FSH signalling consists in physiologically based pharmacokinetic (PBPK) models based on ordinary differential equations (ODE). The objective of such models is to represent qualitatively and/or quantitatively the encoding of specific signals (steady-states signals, sustained oscillations, pulsatility etc.). Early models of FSH dynamics have focused on the FSH-oestrogen interactions in females (Lamport, 1940, Thompson et al., 1969). In these models, secretion and clearance mechanisms are combined as building blocks, with often linear elimination terms and nonlinear (logistic or Hill-type) functions for hormone production, incorporating physiological knowledge on hormonal interactions. Qualitative descriptions of long time behaviour of hormonal regulation models have been largely developed in the 70s, concomitantly with a larger effort of the mathematical physiology community and important advance in theoretical understanding of dynamical systems (Goldbeter, 2002, Keener and Sneyd, 1998, Smith, 1995).

The complexity of the interactions between FSH, LH, steroids, and other key players all along the HPG axis has led to more involved mathematical formalisms. Cyclic patterns of FSH within the menstrual cycle have particularly received lots of attention. Some authors have used algebraic-integro-differential equations (Bogumil et al., 1972, Shack et al., 1971), where the tonic and surge releases of FSH and LH are modelled in details, and gonadotropin accumulation is governed by steroid-dependent decision function. Later works made use of delay differential equations, where explicit time delays account for the time needed for hormone synthesis, in order to obtain a closed autonomous system, more tractable to stability analysis (bifurcation and existence of stable oscillation patterns are shown)(Clark et al., 2003, Harris and Selgrade, 2014). In addition, inspired from works on the male reproductive system (Keenan et al., 2000), stochastic differential equations can be used to represent finely the GnRH pulse-generator mechanisms, which control in turn FSH synthesis, inside a detailed model of the female menstrual cycle (Reinecke and Deuflhard, 2007). Finally, comparison of dynamical models of circulating hormone concentrations with experimental data has to be based on a rigorous statistical framework, adapted to discrete-time measurements. Challenges include in particular the ability to detect occurrences of physiological relevant signal peaks (Urban et al., 1991) or to recover secretion rates (De Nicolao et al., 1999). Several methods exist either based on heuristics and synthetic models (Vidal et al., 2012), deconvolution-based methods (Carlson et al., 2013), or on ensemble models (Keenan and Veldhuis, 2016, Veldhuis et al., 2008).

As many GPCRs, the ligand-stimulated FSHR triggers G-dependent and G-independent intertwined signalling pathways, forming a complex signalling network [see (Gloaguen et al., 2011) for a detailed review] (**Figure 4**). Briefly, the major G-protein coupling to the FSHR is Gαs, which activates adenylate cyclase, resulting in an increase of intracellular levels of cAMP. FSHR also couples to Gαi, which on the contrary inhibits adenylate cyclase, reducing the levels of cAMP (Crepieux et al., 2001). Finally, FSHR has been shown to couple to Gαq (Escamilla-Hernandez et al., 2008) leading to the increase of IP3 levels. The increase of cAMP resulting from Gαs leads to PKA activation, which in turn activates different activating transcription factors. cAMP also activates EPAC, which results in activation of the ERK cascade. Both PKA and EPAC activate Rap1, which in turn activates Akt (Alam et al., 2004).

FSHR has also been shown to activate different proteins in a ligand-dependent and G-protein-independent manner. Among these partners, APPL1 (Nechamen et al., 2004) and PI3K (Wayne et al., 2007) have been shown to activate Akt. These different functional interactions explain how FSHR is able to activate Akt, through both G-dependent and G-independent mechanism, leading to the fine-tuned regulation of transcription. As already discussed above, FSHR, as most GPCRs, interacts with different GRKs, which phosphorylate the receptor at specific sites. GRK-phosphorylated receptors couple to β-arrestin and are internalized. It has been proposed that the FSHR's fate may differ depending on the type of GRK doing the phosphorylation. After internalization, GRK2/3-phosphorylated receptors are either degraded or recycled, whereas GRK5/6-phosphorylated receptors may be capable of initiating G-independent signalling (Kara et al., 2006). In particular, GRK5/6-phosphorylated FSHR complexed with β-arrestin would be able to recruit the ERK signalling module (Raf, MEK and ERK), resulting in G-independent ERK activation (Cassier et al., 2017). The GRK5/6-phosphorylated FSHR complexed with β-arrestin is also able to activate rpS6, participating in the regulation of protein translation (Wehbi et al., 2010b, Trefier et al., 2017). The regulation of protein translation by FSHR also goes through the activation of p70S6 kinase by both PKA and Akt/mTOR-dependent mechanisms (Lécureuil et al., 2005, Musnier et al., 2009). Finally, through the activation of Src and ADAM17, the FSHR transactivates EGFR. In turn, FSH-activated EGFR activates different signalling pathways, among which ERK activation through Ras and Raf (Wayne et al., 2007), and RB1 inactivation through CDK4 (Yang and Roy, 2006).

In addition to the large number of interactions within the FSH interaction network, a key component of the cellular response comes from the dynamical aspects of the signalling network. For instance, different agonists and/or stimulation patterns may activate the same pathways with a distinct temporal signature, leading to very different cellular responses (Kholodenko, 2006, Kholodenko et al., 2010, Lohse and Hofmann, 2015), or may induce dynamical bias in the signalling (Grundmann and Kostenis, 2017, Klein Herenbrink et al.,

2016, Lane et al., 2017). These complex dynamics motivate the use of dynamical modelling techniques to decipher the decoding mechanisms of the FSHR intra-cellular signalling. A standard framework to model the dynamics of signalling pathways uses ordinary differential equations (Ayoub et al., 2016a, Kholodenko et al., 2012, Linderman, 2009). A biochemical reaction network is constructed and, using the law of mass action, this reaction network leads to an evolution law for the species involved in the network. Iterative refinement of the model according to the experimental data is usually needed to answer specific biological questions. This process involves several technical steps that have been detailed for the FSH-induced β-arrestin recruitment dynamics (Yvinec et al.). These steps include solving numerically the ODE model (Higham, 2008), performing numerical optimization to adhere as closely as possible to experimental data (Raue et al., 2013) and, using model selection tools, to discriminate different functional hypotheses (Kirk et al., 2013). The FSH-induced cAMP pathway has been investigated by (Clément et al., 2001) with careful examination of its steady states and parameter sensibility, and compared to cAMP production in granulosa cells. Dynamic interaction between cAMP and mTOR pathways, leading to the regulation of p70S6 kinase, under either FSH or insulin stimulation, was described using experimental data from primary rat Sertoli cells (Musnier et al., 2009). This approach allowed to access to p70S6 kinase phosphorylation rates and was validated by predictions on pharmacologically perturbed conditions. Modelling only the last transcription and translation steps, Quignot and Bois explored the dynamic of sex steroid under FSH stimulation, using *in vitro* and *in vivo* experimental data on rat granulosa cells (Quignot and Bois, 2013). Interestingly, their dynamical model allows studying the effect of endocrine disrupting chemicals on steroidogenesis. We believe the FSH-induced intracellular modelling approaches will be rapidly growing, as more and more biological knowledge shed light on the complexity of its interaction network. Many important questions remain and are mathematically challenging. For instance, to decipher the impact of the pulsatility of the input signal on the downstream intra-cellular signal (Bhattacharya et al., 2017, Clarke et al., 2002), combination of microfluidic devices and mathematical modelling can be particularly relevant (Sumit et al.,

2017). To draw a parallel, interesting developments have been done recently on exploring the GnRH intra-cellular signalling pathways, and in particular the balance between FSH and LH syntheses and releases in response to particular GnRH pulse patterns (Coss, 2017, Stamatiades and Kaiser, 2017). Several interesting modelling approaches have indeed been suggested to explain possible pulse-decoding mechanisms of the GnRHR signalling network, using either deterministic ODE formalism (Fletcher et al., 2014, Pratap et al., 2017) or information theoretic approach and stochastic modelling based on single-cell measurements (Fletcher et al., 2014, Garner et al., 2016).

Up to now, mathematical models of FSH signalling have been extensively used to represent FSH circulating level at the anatomic scale, giving interesting qualitative and quantitative results on the hormonal rhythms. Yet, these approaches intrinsically lack mechanistic interpretation, as the cellular and intra-cellular scales are not included. Further development of FSH intra-cellular modelling, which will benefit from a larger effort in the GPCR modelling community, will help to fill the gap between detailed knowledge on molecular interactions and global understanding on hormonal dynamics. In turn, this will bring important tools for drug screening and development of innovative approaches in drug discovery for reproductive biology.

10. **Impact on gene regulation.**

FSH directly alters the pattern of genes expressed in somatic cells of the gonads by regulating transcriptional as well as post-transcriptional events at the level of mRNA translation and of the miRNA network. The FSH-induced signalling network also indirectly promotes alterations of chromatin condensation in germ cells. Gaining a comprehensive picture on the FSH-regulated gene expression could provide insights on how gonadal somatic cells communicate with their neighbouring germ cells. This could unravel the bases of some spermatogenetic/ovarian failure and potentially lead to the development of innovative therapies against infertility or of new methods of contraception.

1. Chromatin remodelling

The role of Sertoli cells in providing the physical and nutritional requirements met by germ cells to acquire proper size and spermatogenic ability in the adult is known to be regulated, at least in part, by FSH. The Sertoli cell/ germ cell dialog is also illustrated by the ability of FSH to modulate chromatin remodelling at a crucial phase of spermatogenesis, that is, spermiation. This spermatogenesis stage is characterized by the release of elongated spermatids away from Sertoli cells following complex membrane rearrangements, and is particularly sensitive to hormone regulation. At that stage, most histones are dynamically replaced by transition proteins prior to the appearance of protamines, which are in charge with DNA compaction in haploid cells (Sassone-Corsi, 2002). Importantly, knock-out of the FSHR gene in mice (FORKO mice) causes delayed spermatogenesis (Dierich et al., 1998), because of a lack of histone acetylation and ubiquitinylation that precludes their removal from chromatin (Xing et al., 2003). Similar observations have been done in rats treated with fluphenazine decanoate that reduces circulating levels of gonadotropins (Gill-Sharma et al., 2012). Similarly, the FSH signal could alter the compaction of chromatin in granulosa cells of the ovarian follicle, leading to phosphorylation of histone H3 in the promoter regions of the c-fos, serum/glucocorticoid-inducible serine/threonine protein kinase (SGK) and α-inhibin genes, in a PKA-dependent manner (Salvador et al., 2001).

2. Gene transcription

Over the years, several studies have assessed FSH impact on gene transcription in various models and using continuously improving high throughput approaches (**Table 1**). Seminal transcriptomic analyses have highlighted the effect of FSH on the steady-state expression of

100 to 300 mRNA, as soon as after 2 hours of *in vitro* stimulation of prepubertal rat Sertoli cells (McLean et al., 2002). Thereafter, a clearer knowledge on the *in vivo* role of FSH on gene expression has been gained by passive immunization with neutralizing anti-FSH antidodies in rat (Meachem et al., 2005). Among the hormone-regulated transcripts detected, some well-known FSH-responsive genes had been previously identified by *in vitro* approaches (e.g. steroidogenic acute regulatory, androgen-binding protein, connexin-43, phosphodiesterase 4B, cyclin D1 or insulin-like growth factor binding protein-3). In addition, this study also uncovered 48 new targets of FSH regulation, potentially involved in residual body phagocytosis (SCARB1), gene expression (Foxa2, NCOR1, SMAD3), remodelling of the seminiferous epithelium (Reelin, MCPT7, CSPG 4), or cell signalling (MAPK3K1). Concomitant suppression of FSH by neutralizing antibodies and of testosterone action by Flutamide led to contrasted regulations of functional clusters according to the stage of the seminiferous tubule (O'Donnell et al., 2009). For example, genes of the immune function were enriched at stages IX–XIV, which is relevant to secretion of inflammatory mediators (Syed et al., 1995) that is enhanced by the phagocytic activity of residual bodies by Sertoli cells following spermiation. Similar results were obtained in the hpg mouse model (Abel et al., 2008, Sadate-Ngatchou et al., 2004). These mice lack circulating gonadotropins because of a natural mutation in the GnRH gene (Mason et al., 1986), but remain responsive to exogenously administered FSH (Singh and Handelsman, 1996). Hence, they provide a nice model to analyse the FSH-regulated gene regulation *in vivo*. Interestingly, these data depict a temporally dynamic regulation of gene expression by FSH, with an early transcriptional response in Sertoli cells and a more delayed indirect response in germ cells (Abel et al., 2008).

These initial data have been obtained from DNA microarray experiments, providing a rather limited number of differentially expressed genes (DEG). With the advent of next generation sequencing (NGS) technologies, the scientific community might gain a more detailed picture of FSH-regulated genes in physiological as well as pathological conditions, including the identification of splice variants, some of which being known as important mediators of FSH biological action (Foulkes et al., 1993). However, RNAseq analysis of the FSH transcriptome in zebrafish testis disappointedly failed to increase the number of DEG (Crespo et al., 2016).

The transcriptomic analyses presented above have provided an atlas of genes regulated by FSH signalling *in vitro* or *in vivo*, but so far, the transcription factors that could recognize the gene regulatory regions are not known. Systemic examination of the FSH-regulated promoter regions will certainly highlight the involvement of unexpected transcriptional regulators and upstream kinases. For example, the CREB family of transcriptional activators is known to be phosphorylated by PKA in response to FSH, and its functional role in preserving spermatocyte survival has been demonstrated, following injection of a non-phosphorylable version of CREB into the seminiferous tubules (Scobey et al., 2001). However, the requirement of CREB in FSH-regulated transcriptional regulations has to be reconsidered because not so many FSH-responsive genes include cAMP-responsive element (CRE) in their promoter regions (Perlman et al., 2006). For example, a TRANSFAC analysis achieved on transcriptomic data to identify FSH-responsive genes in human granulosa cells revealed an unexpectedly high number of transcripts putatively responsive to the GATA family of transcription factors, whereas CREB-dependent genes were less represented in this study (Perlman et al., 2006). Recent work has revealed that Foxo1 positively and negatively controls the expression of most FSH-responsive genes in granulosa cells (Herndon et al., 2016).

PKA is considered as the master regulator of CREB. However, in FSH-stimulated Sertoli cells, PKA regulates the activity of many other nuclear targets, such as the retinoic acid receptor α (Santos and Kim, 2010), a well-known regulator of germ cell development in the testis. More generally, FSH-induced PKA activity could be involved in global chromatin remodelling, as illustrated above in granulosa cells.

In mammals, the essential role that FSH plays in female fertility is illustrated by the phenotype of FSHβ knockout mice, that exhibit impaired follicular growth, a small uterus and are sterile (Dierich et al., 1998, Kumar et al., 1997). Similar to its role in the seminiferous tubule, in the ovary, FSH indirectly drives oocyte throughout folliculogenesis growth *via* the expression of various factors secreted by follicular cells. Hence, transcriptomic analyses of granulosa cells stimulated by FSH help understanding the dialog between oocytes and somatic cells, and might be useful approaches to optimize assisted reproduction technologies. In that respect, the transcriptomes induced by β-follitropin (Organon), and a glycosylated N-terminal extended FSH variant, FSH1208 (Perlman et al., 2003) in cycling granulosa cells from normal patients undergoing IVF (*i.e.*: male cause of infertility) were compared. Interestingly, this comparison has revealed no difference between both transcriptomes, suggesting that the FSH glycosylation variant could be used in the clinics without apparent modification of the normal gene profile, hence limiting undesirable effects, despite its extended half-life.

Cow has long been used as a model to improve *in vitro* fertilization methods. Gene expression pattern in the cumulus has been associated with the developmental competence of the oocyte it embeds (Assidi et al., 2008, Hamel et al., 2010). In the mature follicle, cumulus cells differentiate from a portion of the granulosa cells and, despite this common origin, cumulus cells and granulosa cells exhibit distinct transcriptomic signatures, which likely underlie different functional properties (Wigglesworth et al., 2015). These differences encompass genes involved in cumulus expansion, steroidogenesis, cell metabolism and oocyte competence (Khan et al., 2015). Recently, the potential role for WNT signalling in potentiating FSH action during dominant follicle selection in cow has been highlighted by RNAseq (Gupta et al., 2014).

Another interesting model is the coho salmon (Oncorhynchus kisutch) because this species only spawns once and then dies. Hence, this is one of a few animal models where follicular cells are synchronous and homogenous, and it can be used to decipher the sequential events that rhythms folliculogenesis. Upon FSH stimulation, it appeared that the genes expressed during the primary growth were associated with cell proliferation and survival, an expression that still markedly increased at the transition to secondary oocyte growth. Then, in vitellogenic follicles, genes associated with cell growth differentiation, growth factor signalling, steroidogenesis and extracellular matrix components were highly expressed. Finally, in maturing follicles, the expression of genes associated with ECM function, growth factor signalling and steroidogenesis-related genes was increased by FSH (Guzmán et al., 2014).

3. mRNA translation

FSHR has a very important trophic role in highly specialized post-mitotic cells, in which it stimulates the production of new constituents. The role of RFSH in the translational regulation of genes has been proposed for the first time in granulosa cells (Alam et al., 2004). During the proliferation of granulosa cells, FSH stimulates the mTOR pathway leading to the expression of cyclin D2, whereas testosterone reduces the effects of FSH by blocking the activity of PKA. Inhibition of ERK results in decreased phosphorylation of an FSH-mediated effector of the mTOR, TSC2 (tuberin) and p70S6K pathway (Alam et al., 2004, Kayampilly and Menon, 2004, Kayampilly and Menon, 2007). FSH stimulates the translation of hypoxia-inducible factor-1 (HIF-1) a transcription factor that regulates the expression of vascular endothelial growth factor (VEGF), plasminogen activator (PLAT), and insulin-like growth factor 2 (IGF-2,) through the PI3K/Akt pathway (Alam et al., 2004). Activation of FSH-mediated mTOR pathway is necessary for the induction of some follicular differentiation markers, including the luteinizing hormone receptor (LH), inhibin-α, aromatase P-450, and βII subunit of PKA. Akt phosphorylation leads to the activation of mTOR which initiates translation by phosphorylating p70S6K on its Thr389 residue (Alam et al., 2004). Finally, activated p70S6K kinase phosphorylates its substrate rpS6 and forms a translation module.

In the differentiating Sertoli cell, the cAMP and PI3K/mTOR pathways contribute to the activation of p70S6K by inducing phosphorylation profiles that are distinct from the model established by Dennis et al. in the case of insulin (Lécureuil et al., 2005, Dennis et al., 1998) (**Figure 5**). Indeed, in the case of insulin-mediated activation, p70S6K is phosphorylated on Thr421 Ser424 and Thr389 (Figure) whereas in the case of FSHR-mediated induction, the activation profile of p70S6K protein depends on the development stage of Sertoli cells: i) during the proliferative phase, FSH induces p70S6K phosphorylation on Thr389 in a PI3K/mTOR-dependent manner and the dephosphorylation on Thr421 and Ser424 in a PKA-dependent manner (Lécureuil et al., 2005, Musnier et al., 2009); ii) during the differentiation phase, FSH activates p70S6K through phosphorylation of Thr389 in a PI3K/mTOR-independent manner (Musnier et al., 2009). Indeed, during the differentiation phase, Akt is phosphorylated directly by the FSHR signalling and not *via* transactivation of the insulin-like growth factor (IGF) 1 receptor as shown in the mitotic phase (Meroni et al., 2004). FSH could use the same transactivation mechanism to trigger the activation of ERK and PI3K, the latter modulating the production of lactate and transferrin, which are essential for the maintenance of germline metabolism (Meroni et al., 2002, Crepieux et al., 2001, McDonald et al., 2006). A recent *in vivo* study demonstrates that INS-R and IGF-1R may be required for FSH-mediated proliferation of Sertoli cells (Pitetti et al., 2013).

From a functional point of view, these FSH-dependent signalling mechanisms may promote changes in the phosphorylation of initiation and elongation factors from the translational machinery present at the 5'UTR of mRNAs. In Sertoli cells, FSH stimulates mTOR activity that, regulates eIF4G phosphorylation and, likely via p70S6K, the phosphorylation of eIF4B, a cofactor of the eIF4A RNA helicase. All these molecular events occurring at the m7GTP 5'cap of mRNAs induce rapid translation (within a few minutes) of mRNAs such as VEGF and c-fos, in the absence of any significant effect on transcription (Musnier et al., 2012). More recently, the p70S6K/rpS6 translation module has been described to form a constitutive molecular assembly with β-arrestins. The G protein-dependent FSH signalling and the β-arrestin-dependent signalling cooperatively contribute to the activation of p70S6K in the β-arrestin/p70S6K/rpS6 complex (Trefier et al., 2017). This complex controls the translation of

5'TOP mRNAs, a subset of mRNAs that contain an oligopyrimidine tract at their 5′ untranslated region. 5'TOP mRNA represent almost 20% of cellular mRNA abundance and encode for ribosomal proteins, poly(A)-binding protein and factors of the translational machinery (Hamilton et al., 2006, Iadevaia et al., 2008, Meyuhas, 2000). The translatome of the FSHR, that is the complete list of target mRNAs translated under the influence of FSH, is not yet established. In the future, such high throughput approaches should help better understanding the dynamics and functional consequence of FSH-mediated translation.

4. miRNA network regulation

In the recent past, accumulating evidence have pointed to microRNAs (miRNAs) as key players in the control of male reproduction (Papaioannou et al., 2011, Papaioannou et al., 2009). This assumption originates from the knock-out of the Dicer gene that is responsible for miRNA biogenesis and processing. Hence, cell responsiveness to FSH extends far beyond the regulation of mRNA transcription and translation, and it is conceivable that a complex microRNA network will respond to the hormone, to regulate the stability, hence dynamics, of various component of the FSH signalling network. Consistently, a rat model where FSH and testosterone action was suppressed *in vivo*, was probed with a miRNA microarray at spermiation (Nicholls et al., 2011). Four of the miRNAs that came out from this analysis were complementary to the PTEN mRNA that localizes in the apical region of the cells, close to mature spermatids. The hormonal input would lead to the degradation of these miRNAs, hence stabilizing PTEN at spermiation. Interestingly, the PTEN protein level is massively enhanced following FSH cell stimulation *in vitro*, leading Sertoli cells to achieve terminal differentiation (Dupont et al., 2010). Since FSH enhances PTEN protein level within minutes, the mechanisms involved is consistent with rapid hormone-induced degradation of a miRNA. miRNA networks might regulate the compartimentalization of FSH signalling components and might control the kinetics of these biochemical reactions, as shown with the PTEN mRNA.

Hence spatial restriction of signalling effectors upon Sertoli cell morphological changes could correlate to differentiation.

In female also, Dicer gene knock-out leads to inefficient ovulation (Gonzalez and Behringer, 2009, Hong et al., 2008, Lei et al., 2010, Nagaraja et al., 2008). In fact, a wide set of miRNA are expressed in granulosa cells, and several of them are responsive to FSH, notably in primary, secondary and antral follicles (Sirotkin et al., 2014, Yao et al., 2009, Yao et al., 2010, Zhang et al., 2017), and appear to play a key regulatory function during follicular growth. For example, the miR10 family is implied in a pro-apoptotic negative feedback loop of TGFβ activity, and its expression is counteracted by FSH (Jiajie et al., 2017). In addition, the FSH signal down-regulates miRNA that prevent progesterone secretion (Yao et al., 2010).

### 11. Impact on proliferation/apoptosis/cell survival.

FSH is an important contributor to the fate of somatic cells of the male and female gonad. Respectively in Sertoli cells and granulosa cells, the hormone regulates proliferation and commitment to differentiation. In addition, in the ovary, FSH protects granulosa cells from atresia, a degenerative process that leads to the selection of a dominant follicle within a developing cohort. This is the main difference with the role of FSH in Sertoli cells, where apoptosis is negligible. The other difference is that each FSH-dependent proliferation/differentiation transition occurs cyclically in the growing follicle, until depletion of the ovarian reserve. At each cycle, FSH promotes proliferation of granulosa cells from the pre-antral stage and then stimulates the expression of the LH-R. In contrast, in male, this transition is not periodical. In rodents, for example, FSH promotes Sertoli cell proliferation perinatally (Griswold et al., 1975, Griswold et al., 1977, Meachem et al., 2005), until the blood-testis barrier forms at puberty. Hence, once mitoses cease, the ultimate size of the Sertoli cell population is reached. Then, FSH sustains the secretory activity of Sertoli cells that supports spermatogenesis during the whole adult life. Hence, at one point, there is a switch in the action of FSH either in male or in female. Understanding the molecular bases for this decision-making is an exciting challenge that deserves to be addressed in current investigations.

   a. Proliferation.

In Sertoli cells, the mitogenic action of FSH is potentiated by several paracrine factors, such as GDNF (Hu et al., 1999) or IGF1 (Pitetti et al., 2013) whose production may vary locally. In granulosa cells, TGFβ family are co-activators of FSH mitogenic effect (Miró and Hillier, 1996). One convincing driver of the proliferation to differentiation switch might rely on the expression of a cytokine-like FSHR receptor with a single transmembrane pass, following alternative splicing of the FSHR gene (Babu et al., 1999). This receptor is expressed early in life and promotes Sertoli cell proliferation upon activation of ERK MAP kinases (Babu et al., 2000). Then, after the switch, the *bona fide* 7TMD FSHR would be expressed and stimulate the cAMP pathway. So far, this cytokine-like FSHR receptor has been identified in granulosa cells only. But strikingly, its action is consistent with the observation that FSH activates ERK MAP kinases in the neonate, to enhance expression of the D1 cyclin that promotes progression through the G1 phase of the cell cycle. In the follicle, enhanced expression of Cyclin D2 (Sicinski et al., 1996) and Cdk4 (Yang and Roy, 2004) have been involved in FSH mitogenic action. A non-mutually exclusive alternative to explain the switch relies on developmental alteration of the miRNA network that would affect the pattern of FSH regulated genes, as suggested by increasing data in granulosa cells (Jiajie et al., 2017, Lei et al., 2010, León et al., 2013, Sirotkin et al., 2014, Yao et al., 2009, Zhang et al., 2017).

   b. Differentiation.

In Sertoli cells, decision-making would occur around 13 days post-partum in rat, while the hormone signal switches to produce cAMP-dependent signalling (Crepieux et al., 2001).

Several paracrine/ endocrine signals antagonize the mitogenic signal elicited by FSH in Sertoli cells (Gallay et al., 2014), the most extensively described being thyroid hormones (Fumel et al., 2012, Holsberger et al., 2005, Quignodon et al., 2007, Van Haaster et al., 1992). The nurturer role of Sertoli cells is nicely illustrated by their ability to stimulate carbohydrate metabolism. More precisely, they convert most of their glucose to lactate, that germ cells use as primary source of energy. This function is stimulated by FSH that accelerates glucose transport and lactate dehydrogenase expression (Galardo et al., 2008). In the follicle, the FSH-dependent expression of the LHR predominates in the second half of the follicular phase, to drive the follicle to the pre-ovulatory, steroidogenic phenotype. Hence, the LHR can be considered as the most prominent granulosa cell differentiation marker induced by FSH. The steroidogenic action of LH is itself potentialized by FSH in human granulosa cells (Casarini et al., 2016c). The mitogenic effect of FSH is counteracted by cyclin inhibitors such as p27kip1 (Robker and Richards, 1998) and by Foxo1 that represses Cyclin D2 expression.

c. Apoptosis and survival.

One major function of FSH is to counteract pro-apoptotic signals in the dominant follicle (Chun et al., 1996). The molecular mechanisms involved include the repression of pro-apoptotic signals such as Bim, whose expression is regulated by Foxo-1 in a PI3 kinase-dependent manner (Shen et al., 2014, Wang et al., 2012). Conversely, FSH also favors the activation of anti-apoptotic signals such as the XIAP family regulated by NF-$\kappa$B in a PI3 kinase-dependent but IKK-independent manner (Li et al., 1998, Wang et al., 2002). β-arrestins might mediate the protective role of FSH, as shown recently in a granulosa cell like (Casarini et al., 2016b).

**12. Biased signalling.**

It is now well established that GPCRs adopt multiple inactive and active conformations that are connected to distinct transduction mechanisms. The notion of signalling bias is coming from this complexity. Indeed, a given ligand or receptor mutation can modify the stabilized conformation of the receptor-ligand complex, as compared to the wild-type receptor-reference-ligand complex (Galandrin et al., 2007, Granier et al., 2007, Kahsai et al., 2011, Kenakin, 2005, Kobilka, 2011, Nygaard et al., 2013, Reiter et al., 2012, Violin and Lefkowitz, 2007, Wacker et al., 2013, Yao et al., 2006, Zürn et al., 2009). Practically, this conceptual framework delineates three levels of possible bias. First, ligand bias that leads to an imbalance between the signalling pathways activated when compared to the reference agonist. Such ligands, by stabilizing a subset of the receptor conformations, potentially lead to drugs that are associated with fewer side effects (Whalen et al., 2011). Second, bias can occur in mutated receptors in which the different pathways normally triggered may be imbalanced or impaired compared to the wild-type receptor. Third, bias that are not dependent on the receptor or the ligand but are associated with modifications of the cellular context. This phenomenon, also known as « conditional efficacy », is associated with differential expression levels of receptor interacting partners or of molecules downstream in the signalling pathways, such as receptor activity-modifying proteins (RAMPs), in different cell types (Christopoulos et al., 2003, Kenakin, 2002). Contrary to the two other types of bias, conditional efficacy is revealed only when comparing a GPCR in different cellular contexts (Landomiel et al., 2014, Ulloa-Aguirre et al., 2011).

a. Ligand bias
    i) Orthosteric and allosteric modulators

Over the years, different classes of small molecules capable of binding to and modulating FSHR have been reported (Arey, 2008, Palmer et al., 2005). In most instances, whether these compounds lead to balanced or biased effects compared to FSH alone has not been determined yet. Some of the described small molecule agonists active at the FSHR are interacting with the transmembrane domains, thus exhibiting an allosteric mode of action (Arey, 2008, Arey et al., 2008, Guo et al., 2004, Guo T, 2004, Palmer et al., 2005, Yanofsky et al., 2006). Three negative allosteric modulators (NAMs) have been reported. The first NAM described was ADX61623, which, while increasing the affinity of FSH binding, blocked FSHR-mediated cAMP and progesterone but not oestradiol production in rat granulosa primary cells. *In vivo*, ADX61623 did not affect FSH-induced preovulatory follicle development (Dias et al., 2011). Two other NAMs, ADX68692 and ADX68693, with structural similarities to ADX61623, were subsequently studied and exhibited biased NAM activities on FSHR in rat granulosa primary cells (Dias et al., 2014, Ayoub et al., 2016b). Indeed, while ADX68692 blocked FSHR-promoted cAMP, progesterone and oestradiol production, ADX68693 inhibited cAMP and progesterone with the same efficacy as ADX68692 but did not block oestradiol production. Interestingly, it was found that ADX68692 but not ADX68693 decreased the number of oocytes recovered fPalmerrom the ampullae (Dias et al., 2014). Van Koppen et al. described a positive allosteric modulator (PAM): Org 214444-0 (Van Koppen et al., 2013). This PAM showed nanomolar FSHR agonistic properties and selectivity over the structurally related LHR and TSHR. When co-incubated with FSH, Org 214444-0 increased FSH binding to its receptor and cAMP activation in a concentration-dependent manner. Furthermore, *in vivo*, Org 214444-0 shows oral bioavailability and mimics the action of FSH in a follicular phase rat fertility model. Unexpectedly, a LHR PAM (Org41841) showed an increase in FSHR at the plasma membrane when used at sub-micromolar concentration. This finding suggests that Org41841 behaves as a pharmacoperone (pharmacological chaperone) at the FSHR. Indeed, this drug was able to rescue the expression of A189V FSHR (Janovick et al., 2009).

Two small molecules have been reported to compete with FSH binding on FSHR. The first competitive antagonist, suramin, has been shown to inhibit testosterone production and FSHR signalling (Danesi et al., 1996, Daugherty et al., 1992). Another one has been later described by Arey and colleagues (Arey et al., 2002). This compound possesses the same antagonistic properties than suramin but with much better specificity for FSHR.

More recently, another series of small molecules acting as ago-PAMs were reported (Nataraja et al., 2015, Sriraman et al., 2014, Yu et al., 2014). Indeed, thiazolidinone derivatives have been reported to activate FSHR signalling in CHO cells and oestradiol production in cultured rat granulosa cells (Sriraman et al., 2014). In addition, optimization of substituted benzamides led to more FSHR-selective molecules relative to other closely related GPCRs, such as LHR and TSHR with better pharmacokinetic properties (Yu et al., 2014).

ii) Glycosylation variants

The subunits of FSH contain several N-linked heterogeneous oligosaccharide structures that play a pivotal role in protein folding, oligomerization, quality control, sorting, and transport (Bishop et al., 1994, Helenius and Aebi, 2001, Ulloa-Aguirre et al., 1999) as well as the functional characteristics of the molecule itself. Carbohydrates account for nearly 20–30% of the hormone's mass (Dias and Roey, 2001, Fox et al., 2001). Each primary sequence of the common α-subunit and FSHβ contains two N-linked oligosaccharides (positions Asn52 and Asn78 on FSHα and Asn7 and Asn24 in FSHβ) (Dias et al., 1998, Ulloa-Aguirre et al., 1999).

Removal of the carbohydrate residue at position 78 on the α-subunit significantly increases receptor binding affinity of human FSH. Likewise, carbohydrate at position 52 on the α-subunit was found to play an essential role in signal transduction as its removal resulted in significant decrease in potency. Interestingly, suppression of both carbohydrates on β-subunit led to a better biopotency than removing both carbohydrates on α-subunit (Bishop et al., 1994). Moreover, β-subunit carbohydrates removal impacts on FSHβ assembly with FSHα resulting in loss of hormone activity and secretion *in vivo* (Ulloa-Aguirre et al., 2003, Wang et al., 2016). Another hypoglycosylated FSH (FSH21/18) was 9- to 26-fold more active than fully-glycosylated FSH (tetra-glycosylated FSH24) in binding assays (Bousfield et al., 2014). Another human FSH deglycosylated variant, which possesses only α-subunit oligosaccharides is significantly more bioactive *in vitro* than the tetra-glycosylated form of the hormone (Walton et al., 2001, Bousfield et al., 2007). In contrast, hyperglycosylated FSH showed an increase in ovulated eggs and subsequent *in vitro* embryo development (Trousdale et al., 2009). Collectively, these results suggest that the naturally occurring FSH isoforms may exhibit biased effects at the target cell level (Timossi et al., 2000).

Supporting this view, it has been reported that partially deglycosylated eLH (eLHdg) is a partial agonist at the FSHR (Wehbi et al., 2010b). Interestingly, siRNA-mediated β-arrestin depletion revealed that eLHdg elicited β-arrestin recruitment to FSHR and activated ERK and rpS6 phosphorylation in a β-arrestin-dependent and G$\alpha$s/cAMP-independent manner (Wehbi et al., 2010b).

iii) Antibodies

In some cases, the structural constraints conveyed on the hormone by a specific monoclonal antibody can affect the receptor's activation mechanism, probably by stabilizing distinct conformations and selectively triggering part of the signalling repertoire (Ulloa-Aguirre et al., 2011). Polyclonal anti-peptide antibodies against ovine FSHβ subunit led to a significant enhancement of biological activity *in vivo* in mice (Ferasin et al., 1997). Likewise, the use of a monoclonal antibody against bovine FSH in the same model (i.e. snell dwarf mice) showed an increase in uterine weigh (Glencross et al., 1993). Equine CG binds only to the equine LH receptor (COMBARNOUS et al., 1984, Guillou and Combarnous, 1983), whereas it exhibits pronounced FSH activity in addition to its LH activity in species other than equine (COMBARNOUS et al., 1978, Licht et al., 1979). Numerous studies evaluated the impact of eCG/anti-eCG complexes on gonadotropin bioactivities. They showed that anti-eCG IgG induced either inhibitory or hyper-stimulation on LH and FSH bioactivity (Hervé et al., 2004).

Furthermore, Wehbi et al. discriminated the nature of these complexes on FSH signalling (Wehbi et al., 2010a). Despite different modulatory effects on cAMP response (*i.e.*: two antibodies increase the signal *vs* one that decreases it), all three complexes tested displayed ERK increased in a β-arrestin dependent manner, thereby displaying biased properties (Wehbi et al., 2010a).

b.   Receptor bias

As presented above, thanks to clinical studies, a large number of FSHR mutations and variants have been examined over the years. Some of them have led to paradoxical results. The most thoroughly studied FSHR variant is the N680S, which is in linkage disequilibrium with the variant T307A. This variant is located at the end of the C-terminal tail of the receptor. This variant is found in both men and women with an allelic distribution of 60% asparagine and 40% serine in the normal population and 50/50 in the infertile population. In the case of ovarian hyperstimulation syndrome (OHSS) in ART, women who are homozygous for serine 680 display a greater severity of symptoms than heterozygotes or homozygotes for asparagine (Daelemans et al., 2004). A study has shown kinetic differences in FSH-induced response at both variants: intracellular cAMP production is faster for N680 than for S680 in granulosa cells. Interestingly, the activation of ERK 1/2, CREB, ARGEG and STARD1 genes expression as well as the production of oestradiol are differentially modulated by this polymorphism (Casarini et al., 2014).

Another well-studied genetic alteration identified at the FSHR is the A189V mutation (Aittomäki et al., 1995). This mutation has been classified as a loss of function because of a lack of plasma membrane expression. However, the phenotype in man was not consistent with a complete lack of function. A study has demonstrated that the A189V FSHR is expressed at very low level at the plasma membrane but still functional. Indeed, when the wild-type receptor is expressed at a comparable level, it does not couple to the Gs/cAMP pathway. However, A189V and wild-type FSHR, both expressed at low levels, are able to induce ERK phosphorylation *via* the β-arrestin-dependent pathway (Tranchant et al., 2011). These observations illustrate nicely the concept of conditional efficacy presented earlier. Indeed, when receptor expression decreases, the stoichiometry of its modulating partners (i.e. β-arrestins in this case) changes, affecting the signalling outcome.

Another case is the N431I mutation in the extracellular loop 1 (EL1) that has been found in a man with undetectable circulating FSH but normal spermatogenesis (Casas-González et al., 2012). This mutation leads to a marked decrease in FSH-induced desensitization and internalization. Finally, it is well established that EL2 of FSHR plays a pivotal role in various events downstream FSH stimulation: three naturally mutations P519T (Meduri et al., 2003), M512I (Uchida et al., 2013) and V514A (Desai et al., 2015) have been described in patients presenting fertility disorders. Six FSHR-specific residues in EL2 have been identified to impair internalization of FSH–FSHR complex and reduce FSH-induced cAMP production (Banerjee et al., 2015). Obviously, a lot of work remains in order to re-evaluate the structure-activity relationships of the FSHR in order to identify possible signalling bias.

## 13. Conclusions

Research on FSHR is barely an emerging topic yet; many advances have been achieved over the last few years that open intriguing prospects in terms of pharmacological control of this receptor with potential applications in ART and contraception. Now that the proof of concept has been achieved that biased signalling exists for FSHR, the different classes of small molecule ligands identified for the FSHR, will have to be further characterized with respect to their pharmacological profiles. Are they balanced or biased? The same goes for the naturally occurring heterogeneity in FSH glycan structure that could affect FSHR activated states and selectively activate intracellular signalling pathways. Robust BRET and/or HTRF assays are now available and should allow exploring the multiple dimensions of efficacy at the FSHR in a quantitative and reliable manner. Finally, molecular characterization of FSHR mutants and variants also needs to take into account the multiple dimension of FSHR efficacy. It is also very important that basic research efforts on FSHR signalling and trafficking continue to deliver new mechanisms and concepts, which are essential to fuel innovative approaches in the control of reproduction.

**Figure legends**

**Figure 1:** Key role of FSH in reproduction. FSH is produced in the pituitary under the control of control neuroendocrine signals (i.e.: Kisspeptin and GnRH) secreted in the hypothalamus. In the testis, FSH acts on FSHR-expressing Sertoli cells in order to promote spermatogenesis. In the ovary, FSH stimulates granulosa cells through FSHR activation, which in turn enhances oestrogen production and induces terminal follicular development.

**Figure 2:** Schematic representation of the key structural elements of the FSHR. The leucine rich repeats (LRR) are represented as blue arrows and numbered from 1 to 12. The pivotal α helix from the hinge region is shown as a blue cylinder. Transmembrane helices (TM1 to TM7) and helix 8 are depicted as orange cylinders. Highly conserved ERW and NPxxY motifs as well as the sulphated Y335, which plays a key role in FSH binding affinity, are indicated in red. FSHα- and FSHβ-subunits are depicted by transparent green ovoid. Extracellular (ECL1 to 3) and Intracellular (ICL1 to 3) loops, NH2-terminal extracellular domain (ECD), hormone binding domain (HBD) and hinge region are delineated on the scheme.

**Figure 3:** Hypothetical model of FSHR trafficking and signalling dynamics controlled by β-arrestins.

**Figure 4:** Schematic view of the FSHR-induced signalling network in CellDesigner. G-dependent pathways are in the centre (Gαs, Gαi and Gαq in light blue) and lead to PKA and Akt activation. In addition to G proteins, some other proteins interact with the receptor, such as APPL1, which also regulate Akt activation. GRK and β-arrestin-dependent action lead to the degradation or recycling of the receptor, but also to G protein-independent ERK

activation. EGFR transactivation also impact PI3K and ERK activation. The major effects of this network on transcription (center bottom) and translation (right bottom) are also represented.

**Figure 5:** Proposed mechanism of p70S6K activation by FSH in granulosa cell and Sertoli cell. **(A)** In granulosa cell, cAMP activates PI3K, Akt and mTOR, which leads to the phosphorylation of p70S6K on Thr389. Once activated, p70S6K phosphorylates rpS6, which participates in the stabilization of HIF1 mRNA. HIF1 then stimulates the expression of various mRNAs including VEGF, aromatase, inhibin α-subunit, LH-receptor and cyclin D2. **(B)** Activation of p70S6K by FSH in Sertoli cells involves PKA-dependent dephosphorylation of p70S6K autoinhibitory Thr421/Ser424 sites while the PI3-K/mTOR pathway constitutively phosphorylates Thr389.

**Aknowledgements**

This work was funded by "ARD 2020 Biomédicament" grants from Région Centre. FDP and AT are recipients of a doctoral fellowship from INRA and Région Centre.

# Bibliography.